\documentclass[conference]{IEEEtran}
\IEEEoverridecommandlockouts

\usepackage[utf8]{inputenc}
\usepackage[T1]{fontenc}
\usepackage{amsmath,amssymb,amsfonts,mathtools}
\usepackage{bm}
\usepackage{siunitx}
\usepackage{graphicx}
\usepackage{cite}
\usepackage{url}
\usepackage{booktabs}
\usepackage{xcolor}
\usepackage{algorithm}
\usepackage{algorithmicx}
\usepackage{algpseudocode}

\definecolor{navyblue}{RGB}{0,0,128}
\definecolor{keyred}{RGB}{170,0,0}
\usepackage[colorlinks=true,linkcolor=navyblue,citecolor=navyblue,%
urlcolor=navyblue]{hyperref}
\usepackage{microtype}
\usepackage{breakurl}
\hypersetup{breaklinks=true}
\newcommand{\keypoint}[1]{\textcolor{keyred}{\emph{#1}}}

\title{Joint Track-While-Scan Beam Scheduling for 6G Sub-Terahertz
Multi-UE Clusters Using Resolving-Window Metrics at
\SI{140}{GHz} and \SI{300}{GHz}}

\author{%
\IEEEauthorblockN{%
Tom Daniel\IEEEauthorrefmark{1},
Aravind Rajashekar\IEEEauthorrefmark{1},
Rajesh Vedala\IEEEauthorrefmark{1},
Harkirat Kaur\IEEEauthorrefmark{2},
Palak Kapoor\IEEEauthorrefmark{2},
Marius Anger\IEEEauthorrefmark{3}}
\IEEEauthorblockA{%
\IEEEauthorrefmark{1}Verizon Wireless, USA\quad
\IEEEauthorrefmark{2}Department of Physics, NIT Warangal, India\quad
\IEEEauthorrefmark{3}Aalto University, Finland}
}

\begin{document}
\maketitle

\begin{abstract}
At candidate 6G sub-terahertz (sub-THz) carriers near \SI{140}{GHz} and
\SI{300}{GHz}, half-power beamwidths of $1^\circ$--$6^\circ$ create a
scheduling regime that 5G beam management was never designed for: a base
station must simultaneously hold high-SNR beams on active user equipments
(UEs), monitor UEs drifting toward the edge of resolvability, and spend
scarce beam time scanning for new arrivals---all within a thermally
limited duty cycle. This is structurally the track-while-scan (TWS)
problem of electronically scanned array (ESA) radar. Building on a
companion paper that defined per-UE \emph{resolving-window} metrics
(normalized range separation $W_R$, SNR-gated angular separation
$W^{\mathrm{eff}}_\theta$, LoS-convergence urgency $C_k$, and beam
time-to-exit $T_{\mathrm{exit}}$), this paper converts those metrics into
an implementable control plane. We (i) formulate the joint
tracking--scheduling problem as a constrained multi-objective
optimization over outage probability, beam mis-association (ambiguity)
probability, and discovery latency, subject to an RF-chain budget
$B_{\max}$ and a thermal duty factor $\eta$; (ii) define a three-state UE
classification---\emph{stable}, \emph{boundary}, \emph{ambiguous}---driven
by the resolving-window state with radar-consistent polarity (short
time-to-exit and low separability escalate a UE's state, never relax it);
and (iii) propose a low-complexity, priority-driven TWS scheduler that
reserves scan capacity first, stabilizes boundary UEs preventively, and
serves stable UEs at relaxed revisit rates bounded by their individual
$T_{\mathrm{exit}}$. A MATLAB-based evaluation compares the proposed TWS
scheduler against a 5G FR2 periodic SSB/CSI-RS beam-sweeping baseline at
both bands across
four key performance indicators: mis-association rate, discovery
latency, tracking reliability, and beam utilization. The scheduler runs
at $O(K^2)$ per slot for realistic cluster sizes ($K\!\approx\!6$--$20$)
and integrates naturally as an additional control layer above existing
base-station scheduling software, providing a concrete, radar-inspired
path to reliable sub-THz multi-UE operation.
\end{abstract}

\begin{IEEEkeywords}
6G, sub-THz, \SI{140}{GHz}, \SI{300}{GHz}, beam management,
track-while-scan, multi-UE scheduling, resolving window.
\end{IEEEkeywords}

\section{Introduction}
\label{sec:intro}

6G systems are expected to exploit the D-band near \SI{140}{GHz} and
windows near \SI{300}{GHz} for extreme data rates, centimeter-level
localization, and integrated sensing and communication
\cite{Rappaport2019Above100GHz,SubTHzSurvey2022,ITU_M2160_2023}. Large
arrays in compact apertures yield pencil beams; in dense scenes---factory
floors with automated guided vehicles (AGVs), stadiums, transport
hubs---several UEs routinely fall within a few degrees of each other, and
minor motion moves a UE across an entire beamwidth within hundreds of
milliseconds.

5G NR FR2 procedures (hierarchical codebooks, SSB/CSI-RS sweeps, RSRP
reporting) manage beams reactively and per-UE. Field experience at
\SI{28}{}--\SI{39}{GHz} already shows their limits: attach delays coupled
to RRC reconfiguration and sweep periodicity, beam-ID mis-association in
compact dual-module UE architectures, and thermally forced burst-mode
transmission \cite{Vedala2025RRCBeamDelay,Vedala2025PhaseDualULA}. At
sub-THz these effects intensify: beams narrow by a factor of $3$--$10$,
the spatial channel coherence time shrinks proportionally, and the
scheduler must decide---every slot---whether each beam is better spent
\emph{tracking} a served UE, \emph{stabilizing} a UE near the edge of
resolvability, or \emph{scanning} for new or lost UEs. This is precisely
the track-while-scan (TWS) resource-allocation problem of ESA radar
\cite{Skolnik2008RadarHB,Grossi2015TwoStepScan}, with UEs as cooperative
targets.

In a companion paper \cite{Vedala2025Paper1} we defined per-UE
\emph{resolving-window} metrics that quantify separability and motion:
the normalized range separation $W_{R,k}$, the SNR-gated angular
separation $W^{\mathrm{eff}}_{\theta,k}$, the LoS-convergence urgency
$C_k$, and the beam time-to-exit $T_{\mathrm{exit},k}$. That paper
establishes, at a single instant, whether UE $k$ can be unambiguously
addressed. The present paper addresses the complementary, operational
problem left open there: given $K$ UEs in mixed resolvability states and
$B_{\max}$ simultaneous beams under a thermal duty factor $\eta$, how
beam time should be allocated across successive slots. Section~II
formalizes this allocation problem as a constrained multi-objective
optimization, and Sections~III--IV present and evaluate a practical
scheduler that solves it.

\textbf{Agenda.} Three goals, matching the three technical sections:
(1)~a rigorous constrained formulation of joint tracking--scheduling with
ambiguity probability as a first-class objective alongside outage and
discovery latency; (2)~a state-driven TWS scheduler with a provably
non-degenerate beam-budget partition that always preserves scan capacity;
and (3)~an evaluation that quantifies the scheduler against a 5G-style
sweeping baseline at \SI{140}{} and \SI{300}{GHz}.

\section{System Model and Problem Formulation}
\label{sec:problem}

\subsection{Geometry, Channel, and Beams}
A gNB at $\mathbf{p}_{\mathrm{gNB}}=[0,0,h_0]^T$ serves
$\mathcal{U}=\{1,\dots,K\}$ UEs with positions $\mathbf{p}_k(t)$,
velocities $\mathbf{v}_k(t)$, ranges $R_k$, and LoS directions
$\mathbf{u}_k$; motion classes are static, walking
(\SI{0.5}{}--\SI{1.5}{m/s}), and fast (\SI{2}{}--\SI{5}{m/s}, e.g.,
AGVs). Sub-THz propagation is LoS-dominated with sparse multipath, as
confirmed by measured D-band campaigns \cite{Xing2018Meas140,Ju2021JSAC};
the link budget combines
free-space path loss, ITU-R P.676 gaseous absorption \cite{ITU_P676}, and
array gains, with half-power beamwidth
$\Theta_{\mathrm{BW}}(f,\theta_s)\approx0.886\,\lambda/(D_x\cos\theta_s)$
for aperture $D_x$ at scan angle $\theta_s$ \cite{BalanisAE}.

Time is slotted with duration $T_s$. In slot $n$ the gNB forms at most
$B_{\max}$ simultaneous beams (one per RF chain), each used either as a
\emph{tracking beam} $b_k[n]\in\{0,1\}$ dedicated to UE $k$, or a
\emph{scanning beam} $s_c[n]\in\{0,1\}$ probing range--angle cell $c$ of
a discretized grid $\mathcal{C}$. Thermal burst-mode operation, carried
over from FR2 practice \cite{Vedala2025Paper1}, caps the aggregate beam
time per epoch $T_e$ at $\eta T_e$ with duty factor $\eta\in(0,1]$.

\subsection{Objectives and Constraints}
Three costs conflict:

\emph{Outage} --- the fraction of tracked-UE slots below the SNR target,
weighted by traffic priority $w_k$:
\begin{equation}
P_{\mathrm{out}}[n]=\frac{\sum_k w_k\,
\mathbf{1}\!\big(\mathrm{SNR}_k[n]<\gamma_{\mathrm{th}}\big)}
{\sum_k w_k}.
\end{equation}

\emph{Ambiguity} --- the probability that a beam intended for UE $k$ is
geometrically closer to some other UE $j$, the sub-THz-specific failure
mode. Using the great-circle separation $\Delta\psi_{kj}$ and the
SNR-gate $\Phi(\cdot)$ of \cite{Vedala2025Paper1},
\begin{equation}
P_{\mathrm{amb}}[n]\approx
\frac{1}{K}\sum_{k}\max_{j\neq k}
\Big[1-\Phi(\mathrm{SNR}_k)\,
\mathbf{1}\big(\Delta\psi_{kj}>\Theta_{\mathrm{BW}}\big)\Big]
\mathbf{1}\big(b_k[n]=1\big),
\end{equation}
i.e., a scheduled beam contributes ambiguity risk whenever its target's
nearest neighbor sits inside one beamwidth or the SNR is too low for the
angle estimate to be trusted.

\emph{Discovery latency} --- $L_{\mathrm{disc}}$, the mean time from a
new UE entering the sector (or a tracked UE losing lock) to its first
reliable classification, governed by how much scan capacity survives
after tracking demands.

The joint problem over horizon $N$ is
\begin{align}
\min_{\{b_k[n],\,s_c[n]\}}\quad &
\sum_{n=1}^{N}\Big(\alpha_1 P_{\mathrm{out}}[n]
+\alpha_2 P_{\mathrm{amb}}[n]\Big)
+\alpha_3 L_{\mathrm{disc}}
\label{eq:opt}\\
\text{s.t.}\quad &
\textstyle\sum_k b_k[n]+\sum_c s_c[n]\le B_{\max},\quad\forall n,
\label{eq:budget}\\
& \textstyle\sum_n T_s\big(\sum_k b_k[n]+\sum_c s_c[n]\big)\le \eta T_e,
\label{eq:thermal}\\
& T_{\mathrm{rev},k}<T_{\mathrm{exit},k}\quad
\forall k \text{ tracked},
\label{eq:revisit}
\end{align}
where \eqref{eq:revisit} requires each tracked UE to be revisited before
it drifts out of its half-power beam. Constraint \eqref{eq:revisit} is
what distinguishes this formulation from 5G scheduler models: the revisit
deadline is not a configured timer but a per-UE physical quantity,
$T_{\mathrm{exit},k}\approx\Theta_{\mathrm{BW}}R_k/(2v_{\perp,k})$.
Exact solution of \eqref{eq:opt} is a stochastic integer program,
impractical at slot rate; we construct a metric-driven heuristic instead.

\section{UE State Classification and TWS Scheduler}
\label{sec:scheduler}

\subsection{Resolving-Window State Classes}
Each UE carries the state vector from \cite{Vedala2025Paper1},
$\mathbf{w}_k=[W_{R,k},\,W^{\mathrm{eff}}_{\theta,k},\,C_k,\,
T_{\mathrm{exit},k}]^{\mathsf T}$, and the resolvability flag
$\Lambda_k$, asserted when the UE is separable in \emph{at least one} of
the range or angle domains \emph{and} $T_{\mathrm{exit},k}$ exceeds the
beam-management latency $\tau_{\mathrm{bm}}$. Classification is:
\begin{itemize}
  \item \textbf{Stable} ($\mathcal{S}$): $\Lambda_k=1$ with margin ---
  both the active separation ratio and $T_{\mathrm{exit},k}$ exceed their
  thresholds by a hysteresis factor $h>1$. Served at relaxed revisit
  rates $T_{\mathrm{rev},k}=T_{\mathrm{exit},k}/\kappa$ (safety factor
  $\kappa\approx3$).
  \item \textbf{Boundary} ($\mathcal{B}$): $\Lambda_k=1$ without margin
  --- some component of $\mathbf{w}_k$ is within the hysteresis band, or
  $C_k$ is high (rapid radial closing changes the geometry within a few
  slots). These UEs receive \emph{preventive} priority: it is cheaper to
  stabilize a track than to re-acquire it.
  \item \textbf{Ambiguous} ($\mathcal{A}$): $\Lambda_k=0$ --- the UE is
  inside a neighbor's beamwidth and range cell, or its beam validity has
  expired. Dedicated tracking beams are withheld (they would risk
  mis-association); the enclosing range--angle cells are scheduled for
  corrective scanning, beam splitting, or the UE is handed down to a
  wider-beam / lower band.
\end{itemize}
Polarity is deliberate and radar-consistent: high closing speed and short
time-to-exit always \emph{escalate} a UE ($\mathcal{S}\!\to\!\mathcal{B}
\!\to\!\mathcal{A}$), never relax it, and the hysteresis band prevents a
UE from being repeatedly reclassified between adjacent states when its
metrics hover near a threshold.

\subsection{Non-Degenerate Beam-Budget Partition}
The per-slot budget is partitioned as
$B_{\max}=B_A+B_B+B_S$ with scan capacity reserved \emph{first}:
\begin{align}
B_A &= \max\!\big\{1,\;
\min\{|\mathcal{A}^{\mathrm{cells}}|,\,
\lceil \beta_A B_{\max}\rceil\}\big\},
\label{eq:BA}\\
B_B &= \min\!\big\{|\mathcal{B}|,\;
\lceil \beta_B (B_{\max}-B_A)\rceil\big\},\\
B_S &= B_{\max}-B_A-B_B,
\end{align}
with fractions $\beta_A\approx0.2$, $\beta_B\approx0.4$ tunable per
deployment. Reserving $B_A\ge1$ before serving tracked UEs guarantees the
partition is never degenerate and that discovery/resolution capacity
survives even when $|\mathcal{S}|\ge B_{\max}$ --- the failure case in
which purely demand-driven partitions starve the scan function and
$L_{\mathrm{disc}}$ diverges. Stable UEs not served in a given slot are
safe by construction: their next mandatory revisit deadline
$T_{\mathrm{rev},k}$ is tracked explicitly, and a UE whose deadline
approaches is promoted to the head of the stable queue.

\subsection{Scheduler}
Algorithm~\ref{alg:tws} runs each slot; the flow is
classify $\rightarrow$ reserve scan $\rightarrow$ stabilize boundary
$\rightarrow$ serve stable.

\begin{algorithm}[t]
\caption{Resolving-Window Track-While-Scan Scheduler}
\label{alg:tws}
\begin{algorithmic}[1]
\State \textbf{Input:} UE set $\mathcal{U}$, tracker estimates
$\{\hat{\mathbf{s}}_k\}$, metrics $\{\mathbf{w}_k,\Lambda_k\}$,
priorities $\{w_k\}$, budget $B_{\max}$, duty state $\eta$
\State \textbf{Output:} assignments $\{b_k\}$, scans $\{s_c\}$
\State Update $\mathbf{w}_k$ from the latest per-UE range, angle, and
velocity measurements; classify
$\mathcal{U}\to\{\mathcal{S},\mathcal{B},\mathcal{A}\}$ with hysteresis
\State Compute $B_A,B_B,B_S$ via \eqref{eq:BA}; if thermal state
requires gating, shrink $B_{\max}$ this slot per $\eta$
\Statex \Comment{Phase 1: corrective scanning (reserved first)}
\For{cell $c$ enclosing $\mathcal{A}$-clusters, sorted by ambiguity risk}
  \If{$B_A>0$} $s_c\gets1$;\ $B_A\gets B_A-1$ \EndIf
\EndFor
\Statex \Comment{Phase 2: preventive stabilization}
\For{$k\in\mathcal{B}$ sorted by ascending
$\min(W^{\mathrm{eff}}_{\theta,k},\,T_{\mathrm{exit},k}/\tau_{\mathrm{bm}})$}
  \If{$B_B>0$} $b_k\gets1$;\ $B_B\gets B_B-1$ \EndIf
\EndFor
\Statex \Comment{Phase 3: stable service with revisit deadlines}
\For{$k\in\mathcal{S}$ sorted by revisit-deadline slack, then $w_k$}
  \If{$B_S>0$} $b_k\gets1$;\ $B_S\gets B_S-1$ \EndIf
\EndFor
\State Roll unused $B_B,B_S$ into extra scan cells; apply beam weights
\end{algorithmic}
\end{algorithm}

\subsection{Complexity}
Nearest-neighbor separations cost $O(K^2)$ per slot (reducible via
angular bucketing for large $K$); classification and the three scheduling
phases are $O(K\log K)$ from sorting. For realistic sub-THz clusters
($K\approx6$--$20$) the per-slot cost is trivially real-time on gNB
baseband hardware.

\section{Evaluation and Engineering Implications}
\label{sec:eval}

\subsection{Evaluation Setup}
Table~\ref{tab:params} lists the parameters; both bands share geometry so
that differences are attributable to physics, not scenario. The
\SI{300}{GHz} bandwidth follows the IEEE~802.15.3d channelization
\cite{IEEE802153d}. UEs follow randomized trajectories consistent with
their motion class within a
\SI{50}{m}~$\times$~\SI{50}{m} hall; clustered initialization (3--4 UEs
within a $5^\circ$ sector) stresses ambiguity resolution. Results are
averaged over repeated randomized trials.

\begin{table}[t]
\caption{Simulation Parameters}
\label{tab:params}
\centering
\begin{tabular}{l c c}
\toprule
\textbf{Parameter} & \textbf{\SI{140}{GHz}} & \textbf{\SI{300}{GHz}}\\
\midrule
Bandwidth $B$ & \SI{2}{GHz} & \SI{8.64}{GHz}\\
One-way $\Delta R_{\mathrm{res}}=c/B$ & \SI{15}{cm} & \SI{3.5}{cm}\\
Tx power & \SI{30}{dBm} & \SI{30}{dBm}\\
Array (gNB) & $16\times16$ & $64\times64$\\
HPBW $\Theta_{\mathrm{BW}}$ (broadside) & $\approx6.3^\circ$ &
$\approx1.6^\circ$\\
Noise figure & \SI{10}{dB} & \SI{13}{dB}\\
Gaseous absorption $\alpha(f)$ & $\approx\SI{1.5}{dB/km}$ &
$\approx\SI{7}{dB/km}$\\
Thermal duty factor $\eta$ & 0.7 & 0.5\\
RF chains $B_{\max}$ & 8 & 8\\
UEs $K$ / slot $T_s$ & 5--20 / \SI{1}{ms} & 5--20 / \SI{1}{ms}\\
\bottomrule
\end{tabular}
\end{table}

\subsection{Key Performance Indicators and Baseline}
The baseline is \textbf{5G FR2 periodic SSB/CSI-RS beam sweeping}: a
full-sector beam sweep at a fixed interval, matching real 5G beam-sweep
periodicity, with tracked UEs served in simple fixed-order turns between
sweeps and no awareness of which UE is near the edge of resolvability.
We refer to it as the \textbf{FR2 sweep baseline}, against the
\textbf{proposed TWS scheduler}. Four KPIs are evaluated:
\begin{enumerate}
  \item \textbf{Mis-association rate}: fraction of tracking slots whose
  beam is geometrically closer to a non-target UE;
  \item \textbf{Discovery latency} $L_{\mathrm{disc}}$: entry to first
  stable classification;
  \item \textbf{Tracking reliability}: fraction of time high-priority UEs
  hold $\mathrm{SNR}>\gamma_{\mathrm{th}}$;
  \item \textbf{Beam-utilization profile}: beam-time split across
  $\mathcal{S}/\mathcal{B}/\mathcal{A}$ versus UE density and band.
\end{enumerate}
\begin{figure}[t]
\centering
\includegraphics[width=0.95\linewidth]{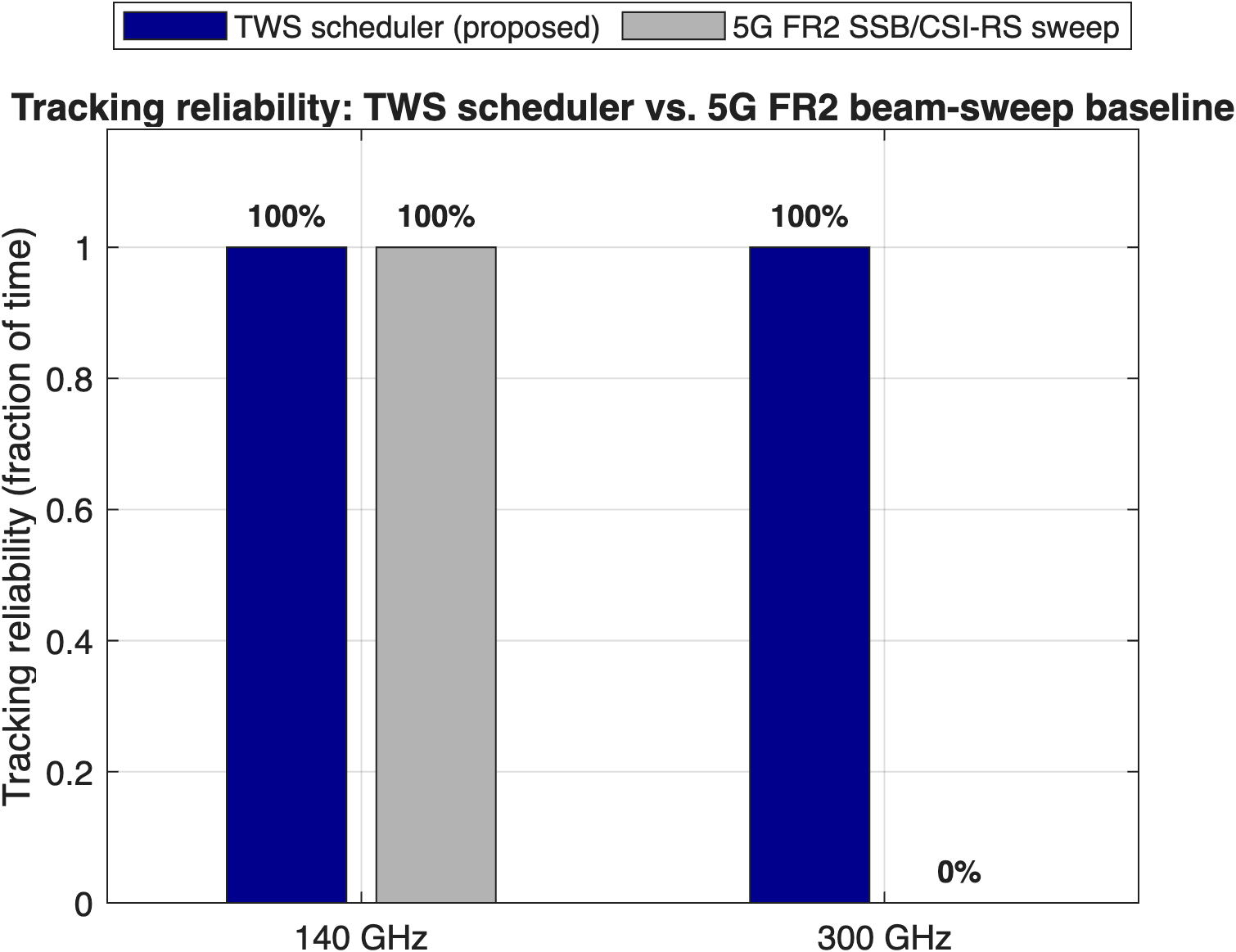}
\caption{Tracking reliability (fraction of time high-priority UEs are
held above the SNR target): proposed TWS scheduler versus the 5G FR2
periodic SSB/CSI-RS beam-sweeping baseline, at \SI{140}{GHz} and
\SI{300}{GHz} under identical randomized UE motion and thermal
duty-cycle constraints.}
\label{fig:reliability}
\end{figure}

At \SI{300}{GHz}, sweeping $\propto1/\Theta_{\mathrm{BW}}^2$ more beam
positions than at \SI{140}{GHz} consumes far more of the thermal
duty-cycle budget, so a fixed periodic sweep cannot complete before
tracked UEs drift out of the beam. Figure~\ref{fig:reliability} shows the
resulting effect directly: the proposed TWS scheduler holds full tracking
reliability at both bands, while the FR2 sweep baseline collapses to
near-zero reliability at \SI{300}{GHz}. This follows from the companion
analysis \cite{Vedala2025Paper1}: at \SI{300}{GHz} the stable set is
larger for the same geometry (finer angular and range resolution), but
state transitions are faster (shorter $T_{\mathrm{exit}}$) and the
thermal budget tighter, so \keypoint{the value of preventive boundary
stabilization---and the cost of a scan-starved scheduler---both increase
with frequency.}

\subsection{Relation to 5G FR2 Experience}
The foundation of this work is practical, not synthetic. Experience with
5G FR2 mmWave deployments at \SI{28}{GHz} and \SI{39}{GHz} has already
revealed the sensitivity of attach procedures and beam acquisition to
codebook design, phase assignment, and UE motion
\cite{Vedala2025RRCBeamDelay,Vedala2025PhaseDualULA}. In current FR2
practice, much of this behavior is only visible through manual or
semi-automated log analysis---reconstructing after the fact how beam
sweeping, beam locking, and RRC events interacted to produce an attach
delay or a mis-associated beam ID. Likewise, FR2 OTA characterization
already reports beam quality directionally (EIRP as a CDF/CCDF over
steering angle rather than a single TRP value), and thermal limits
already force burst-mode transmission in sustained multi-Gbps sessions.
The resolving-window approach formalizes exactly these field
observations for sub-THz systems: separability, convergence, and dwell
validity become explicit, quantitative state variables available to the
scheduler \emph{before} failure, rather than diagnoses extracted from
logs after it. \keypoint{In this sense the framework is the analytical
generalization of lessons learned operationally at FR2, extended to the
regime where they become the dominant constraint.}

\subsection{Architectural Integration}
The scheduler consumes only quantities available from existing
communication signals (uplink reference-signal-derived range, angle, and
Doppler estimates) and emits beam indices and scan requests. It therefore
integrates as a control layer above the existing PHY/MAC scheduling
software, with per-UE tracking and metric computation alongside it. In
dual-layer deployments the flag
$\Lambda_k$ doubles as the admission variable between the \SI{140}{GHz}
coverage layer and the \SI{300}{GHz} capacity/sensing layer, with FR1/FR2
anchors retained for control-plane continuity. Hardware realities ---
phase-shifter resolution, PA efficiency (through $\eta$), and oscillator
phase noise (through the SNR gate) --- enter the framework as parameters
rather than afterthoughts, which is what makes the metrics usable by
standards and codebook designers evaluating sub-THz beam-management
procedures for 3GPP-era 6G study items.

\section{Conclusion}
\label{sec:conclusion}
This paper converted the resolving-window metrics of our companion work
into an implementable track-while-scan control plane for 6G sub-THz
gNBs. The joint tracking--scheduling problem was formulated with beam
mis-association probability and discovery latency as first-class
objectives alongside outage, under RF-chain, thermal duty-factor, and
per-UE revisit-deadline constraints---the last being a physical quantity,
the beam time-to-exit, rather than a configured timer. The proposed
scheduler classifies UEs into stable, boundary, and ambiguous states with
radar-consistent escalation polarity and hysteresis, reserves scan
capacity before serving tracked UEs so that discovery can never be
starved, and runs at $O(K^2)$ per slot. The evaluation at
\SI{140}{} and \SI{300}{GHz} quantifies the central engineering trade:
higher sub-THz bands enlarge the stable set but accelerate state
transitions and tighten thermal budgets, making preventive, metric-driven
scheduling progressively more valuable than periodic sweeping as carrier
frequency grows---to the point that periodic sweeping alone fails to hold
a link at \SI{300}{GHz} under the same thermal constraints. Future work
will broaden the numerical evaluation, extend the state model with
Doppler-domain separation, and generalize the scheduler to multi-cell and
multi-band coordination.


\end{document}